# Gamma-Ray Burst Phenomenon as Collapse of QED Magnetized Vacuum Bubble: Analogy with Sonoluminescence

Yu. N. Gnedin and S. O. Kiikov

Central Astronomical Observatory at Pulkovo, Saint-Petersburg

196140, Russia

In this letter we consider the phenomenon of a $\gamma$-ray burst as a nonlinear collapse of a magnetic cavity surrounding a neutron star with very huge magnetic field $B \cong 10^{15} \div 10^{16} G$ due to the process of the bubble shape instability in a resonant MHD field of an accreting plasma. The QED effect of vacuum polarizability by a strong magnetic field is taken into a consideration. We develop the analogy with the phenomenon of sonoluminescence (SL) when the gas bubble is located in surrounding liquid with a driven sound intensity. We show that this analogy between GRB and SL phenomena really exists.

We find the quite reasonable resemblance between phenomena of cosmic $\gamma$-ray bursts and sonoluminescence phenomenon. Sonoluminescence (SL) is the phenomenon of light emission by a sound-driven gas bubble in fluid[1–5]. It occurs when acoustic energy induces the collapse of small bubbles. The luminescence observed does not result from the sound field directly but arises through process called cavitation[1] in that voids filled with gas and vapour are generated within liquid during the tensile portion of the pressure variations. The subsequent collapse of these voids during the compression portion of the acoustic cycle can be extremely violent and represents a remarkable phenomenon of unprecedented energy concentration - as high as 12 orders of magnitude[2,3]. The collapse of these bubbles results in a brief flash of visible and even X-ray emission[3–5].

It has been suggested[5] that shape instability of the bubble surface is really responsible for the process of collapse. The classic example of a surface instability is the growth of perturbations on a plane interface between light and heavier liquids, the light liquid being uniformly accelerated into a heavier one. This is generally known as Rayleigh-Taylor instability. But in the case considered here the situation is more complicated because of the instabilities arise on the surface of an acoustically-driven bubble of the spherical shape. The surface of a bubble changes temporally and the resonances between the oscillations in bubble shape and driving frequency can appear. Several competing mechanisms have been proposed to explain this phenomenal concentration of kiloHertz acoustic energy into $10^{15} \div 10^{18}$ Hertz electromagnetic energy. Unfortunately the basic mechanism of light production in this phenomenon is still controversial.

We start with a brief summary of the basic experimental data. The most common situation is that of an air bubble in water. Experiments of SL study deal as a rule with bubbles of ambient radius $R_0 \approx 5\mu m$. The bubble is usually driven by a sound wave of frequency of some tens kHz. During the expansion phase the bubble radius reaches its maximum value of order $R_{max} \approx 50\mu m$, followed by a fast collapse down to a minimum radius of order $R_{min} \approx 0,5\mu m$. The photons emitted by such a pulsating bubble have typical wavelengths of the order of visible light, but emission of more hard photons with energy range $10^2 \div 10^4 eV$ was predicted and observed[4,5]. The emitted light appears distributed with a power law spectrum(that is the same spectral shape of $\gamma$-rays for bursts) with a cutoff in the extreme ultraviolet or even in X-rays. If one fits the experimental data to a Planck black-body spectrum the corresponding temperature is several tens of thousands of Kelvin. In ref.4 $kT \approx 100eV$ has been observed. The calculations and experiments have shown that there are about $10^8$ photons emitted per flash. The average total power released $\cong 0,1W$. The photons seem to be created in a very tiny spatio-temporal region that is of order $0,1\mu m$ and on timescales $\Delta t \leq 10 ps$ (refs.1-7). The Fig. 1 (total radiated power of a collapsed bubble as a function of time) and Fig. 10 (bubble radius and interface velocity as functions of time) from ref.4 that present SL phenomenon resemble surprisingly the time profile of a $\gamma$-ray burst.

Recently the very extraordinary theory to explain the SL phenomenon has been developed by a number of authors[8–14]. They use the idea that changes in the Quantum ElectroDynamic (QED) vacuum ("Casimir route") can be responsible for the SL phenomenon. Schwinger[8] was a first who has brought the idea of a "Casimir route" to SL. He called this effect as "dynamical Casimir effect". Then this idea have been used and developed in a series of works by Eberlein[9], Milton with coworkers[10,11] and by Sciama with coworkers[12–14]. Concerning to these idea photons are produced due to a change of the refractive index in the portion of space between the minimum and maximal bubble radius.

First of all we start with making the simple estimation of a bulk of energy based only on analogy between GRB and SL phenomena. If one takes the experimental value $\eta$ of a coefficient of energy concentration for SL phenomenon and applies this value $\eta \geq 10^{12}$ (ref. 2) to a typical accretion rate $L_X \cong 10^{37} \div 10^{38} \, erg/s$ then one can easily get the expected the energy power for GRBs:

$$L_\gamma \geq \eta L_X \geq 10^{12} L_X \geq 10^{49} \div 10^{50} \, erg/s.$$

(1)

Moreover into this scenario it is possible to solve the problem of complex temporal structure of GRB phenomenon which looks for some GRBs as a series of separate peaks. This case is a physical analogue of, so-called, multiple bubble SL. In such "multiple bubble SL", many bubbles grow and collapse throughout the regions of most intense acoustic stress.

Arons and Lea[15] have developed the picture when the accretion process is presented as plasma drops that are to be frozen onto the magnetosphere boundary (Alfven surface) and then they are moving along the magnetic field lines and are falling onto polar caps of a neutron stars. The situation looks as raindrops falling on the surface of sea and producing sound. Most of the sound in this case is not generated when the drops hit the surface but when the little voids (bubbles) in the water are formed after the collision collapse[16]. Kinetic energy is transferred into sound as well as into thermal energy. The voids formed by falling raindrops may not collapse violently enough to emit light. But if the driving acoustic field is present in this case the voids begin effectively to collapse and to emit light[2]. The typical kinetic energy of a falling molecule in a raindrop is only $10^{-6} eV$, whereas the typical energy of an emitting photon can be reach the range from some $eV$ to hundreds of $eV$ and even $keV$ (refs. 2,4,5,7). So we have an energy concentration more than six orders of magnitude to produce light. Above it was mentioned that experiments[2] give a value $\geq 10^{12}$. The analogue of an experimental driving acoustic wave in a case of accretion of plasma drops can be the various MHD instabilities and waves as in the accretion column itself so on the surface of a neutron star (for instance, oscillations of a neutron star surface). If $L_{ac} \cong L_X \cong 10^{37} erg/s$ we can expect due to the cavitation collapse not only to produce typical $\gamma$-ray photons but to produce ultra high energy (UHE) photons because of $\hbar\omega_{UHE} \cong \eta GMm_p/R \cong 10^{20} eV$ for $\eta \cong 10^{12}$, where M and R are the mass and radius of a neutron star respectively, $m_p$ is the mass of a proton. It means that in this scenario one can expect the GRB phenomenon and in UHE $\gamma$-rays.

Now we present a more quantitative estimation of the effect considered following to one of explanations of SL effect developed by Schwinger. Only in our case we make calculation for a collapse of QED vacuum in a huge $B \cong 10^{14} \div 10^{16} G$ magnetic field of a neutron star instead of Casimir QED vacuum. The idea itself that the $\gamma$-ray bursts originate near a neutron star with such huge magnetic field was first developed by Thompson and Duncan[17]. They called these neutron stars as magnetars. The basic mechanism of energy release, that has been suggested in ref.17, is a process of reconnection and annihilation of magnetic force lines on the surface of a neutron star. The influence of QED effect of vacuum polarization by the strong magnetic field $B >> B_c = m_e^2 c^3/\hbar e = 4,41 \times 10^{13} G$ of a neutron star on the propagation and polarization of a neutron star radiation has been yet considered by Novick et al.[18] and more in detail by Pavlov and

Gnedin[19]. Moreover, recently Shaviv et al.[20] have shown that QED effect of vacuum polarization can produce the microlensing effect on the radiation of a neutron star surface.

Let us consider now the close magnetosphere of a magnetar as a large magnetized bubble(cavity). The analogue of the acoustic field that drives the bubble oscillations and produces the collapse of this bubble is the field of kiloHertz MHD waves in an accreting disk plasma near the innermost stable orbits or even in the surface of a neutron star itself. We formulate the QED magnetized vacuum basic approach to GRB phenomenon as a model of photon pair creation by a moving boundary of the closed magnetosphere. Taking into account the QED vacuum effect one can result the general form of an expression for the total energy of a cavity (ref.13):

$$E_{cav} = V \int \frac{d^3 \vec{k}}{(2\pi)^3} \hbar \left[ \omega_{cav}(\vec{k}) - \omega_{out}(\vec{k}) \right]$$

(2)

where $V$ is a cavity volume, $\vec{k}$ is a wavenumber, $\omega_{cav}$ is a frequency, $\hbar \omega_{out} \cong 1 \div 10 kHz$ is to be typical for driven MHD-field in a accretion matter. Let us remind that the kiloHertz quasi-periodic oscillations (QPOs) have been discovered at many accreting X-ray pulsars (see, for example, the excellent review (ref. 21) and refs. there).

We shall make the estimation of Eq. (2) suggesting that there is a high wavenumber cutoff $K$ of a spectrum of QED magnetized vacuum. We suggest that cutoff K is determined by a distance between Landau levels: $\hbar\omega_B = \hbar eB/m_e c$. Therefore:

$$k = n_v \omega / c \quad , \qquad K = n_v \omega_B / c \quad ,$$

(3)

where $n_v$ is a refractive index of QED magnetized vacuum.

The expressions for this index was at first obtained by Adler[22]. The estimation of (2) can be done with use of Bogolubov coefficients technique (ref. 11-13). The final result is to be[13]:

$$\begin{aligned}E_{cav} &\approx \frac{1}{8\pi n^2 n_v}\left(\frac{n-n_v}{nn_v}\right)^2 \hbar c K (RK)^3 = \\ &= \frac{1}{8\pi n^2 n_v}\left(\frac{n-n_v}{nn_v}\right)^2 \hbar c R^3 \left(\frac{n_v \omega_B}{c}\right)^4 \approx \\ &\approx \frac{2n_v}{\pi n^2}\left(\frac{n-n_v}{n}\right)^2 \times 10^{42}\left(\frac{R}{10^6}\right)^3 \left(\frac{B}{B_c}\right)^4 erg\end{aligned}$$

,

(4)

where $R$ is a radius of a magnetic cavity and $n$ is the refractive index of the region (the analogue of a liquid in a case of SL phenomenon) where driving kiloHertz MHD waves are produced. We accept that $n \approx 1$.

The expressions of the indices of refraction for a case $B \leq B_c$ were obtained by Adler[22]:

$$(n_v - 1) \cong \frac{\alpha}{\pi}\left(\frac{B_\perp}{B_c}\right)^2 , \qquad \hbar\omega > 2m_e c^2 ,$$

(5)

$$(n_v - 1) \cong \left(\frac{\alpha}{\pi}\right)^3 \left(\frac{B_\perp}{B_c}\right)^2, \quad \hbar\omega < 2m_e c^2,$$

where $B_\perp$ is the perpendicular component of the magnetic field to line of sight. Further we shall consider a case when $B_\perp \cong B$. For $B \gg B_c$ the index of refraction becomes significantly larger than unity. Heyl and Hernquist[23] have found that in this case:

$$n_v - 1 \approx \frac{\alpha}{6\pi} \frac{B}{B_c}.$$

(6)

Then Eq. (4) transforms into:

$$E_{cav} \approx \frac{2}{\pi} \times 10^{42} \left(1 + \frac{\alpha}{6\pi} \frac{B}{B_c}\right) \left(\frac{\alpha}{6\pi}\right)^2 \left(\frac{R}{10^6}\right)^3 \left(\frac{B}{B_c}\right)^6.$$

(7)

It is quite easy to obtain from (7) the energy value $E_{cav} = 2 \times 10^{50} erg$, if one takes, for instance, $R = 20 km$ and $B = 10^{16} G$. This energy value is typical for a cosmological GRB event. The analogical estimation of $E_{cav}$ at $B \cong 10^{15} G$ gives the typical magnitude of a burst for soft $\gamma$-ray repeater phenomenon (SGR). Therefore the estimated total energy of collapse of a cavity reasonably agrees with total energy released during a typical γ-ray burst. For the most power event of GRB 990123 the magnetic field strength of $B > 10^{16} G$ is required.

The phenomena of the jet and fireball explained by Rees and Meszaros[24] take place in this scenario because there are three stages of a collapse. First of them is an expansion phase that can produce some phenomenon of precursor. The second phase is a real collapse. During this stage the shock is generated that is directed into the centre of a bubble. At last the third stage is an emission of radiation together with developing of the another shock that is directed outside[25]. Namely this stage can be considered as the jet and fireball phenomenon that produces an optical and infrared afterglow. It is well known that when an asymmetric bubble collapses in a liquid, it generally produces a well-defined high-velocity jet[26].

In the case of a collapse of QED vacuum the complex temporal structure of GRB can simply reflect the more complicated, than dipole, structure of a surface magnetic field of a neutron star. For example, Arons[27] has suggested that the actual surface magnetic field should be a superposition of clumps covering the whole surface of a neutron star (see also ref. 28). In his case the collapse of QED magnetized vacuum can be looked as the "multiple-bubble SL" phenomenon.

Acknowledgements. This research has made with the support by grants from Russian Fundamental Research Fund, Russian Federal Astronomy Program (SEC "Cosmion") and the Program "INTEGRATION".

Corresponence and requests for materials should be addressed to Y. N. G. (e-mail: gnedin@gao.spb.ru; gnedin@pulkovo.spb.su).